\documentclass[twocolumn,showpacs,preprintnumbers,amsmath,amssymb,prb]{revtex4}

\usepackage{graphicx}
\usepackage{dcolumn}
\usepackage{bm}
\usepackage{color}

\begin{document}

\preprint{}

\title{Disorder-sensitive superconductivity in the iron silicide Lu$_2$Fe$_3$Si$_5$ studied by the Lu-site substitutions}

\author{Tadataka Watanabe}
\author{Hiroki Sasame}
\author{Hiroaki Okuyama}
\author{Kouichi Takase}
\author{Yoshiki Takano}
\affiliation{Department of Physics, College of Science and Technology (CST), Nihon University, Chiyoda-ku, Tokyo 101-8308, Japan}
\date{\today}

\begin{abstract}
We studied effect of non-magnetic and magnetic impurities on superconductivity in Lu$_2$Fe$_3$Si$_5$ by small amount substitution of the Lu site, which investigated structural, magnetic, and electrical properties of non-magnetic (Lu$_{1-x}$Sc$_x$)$_2$Fe$_3$Si$_5$, (Lu$_{1-x}$Y$_x$)$_2$Fe$_3$Si$_5$, and magnetic (Lu$_{1-x}$Dy$_x$)$_2$Fe$_3$Si$_5$. The rapid depression of $T_c$ by non-magnetic impurities in accordance with the increase of residual resistivity reveals the strong pair breaking dominated by disorder. We provide compelling evidence for the sign reversal of the superconducting order parameter in Lu$_2$Fe$_3$Si$_5$.
\end{abstract}


\maketitle

Recent discovery of high-$T_c$ superconductivity in the FeAs systems has shed a brilliant light on Fe-based substances as a rich vein of new exotic superconductors.\cite{Kamihara} In addition to deeper studies of the FeAs systems, it is also indispensable to explore the exotic superconductivity in Fe-based substances other than the FeAs family. Ternary iron-silicide Lu$_2$Fe$_3$Si$_5$ is a non-FeAs-family superconductor discovered in 1980.\cite{Braun} This compound crystallizes in the tetragonal Sc$_2$Fe$_3$Si$_5$-type structure consisting of a quasi-one-dimensional iron chain along the $c$ axis and quasi-two-dimensional iron squares parallel to the basal plane.\cite{Braun4} The superconductivity occurs at $T_c$ = 6.0 K which is exceptionally high among the Fe-based compounds other than the FeAs family. According to M$\ddot{o}$ssbauer experiments, Fe atoms in Lu$_2$Fe$_3$Si$_5$ carry no magnetic moment.\cite{Braun2} Taking into account the absence of superconductivity in the isoelectronic Lu$_2$Ru$_3$Si$_5$ and Lu$_2$Os$_3$Si$_5$,\cite{Johnston} Fe 3$d$ electrons in Lu$_2$Fe$_3$Si$_5$ should play significant role in the occurrence of the superconductivity.

To unveil the pairing mechanism of the exotic superconductivity, it is crucial to determine the superconducting gap function. In Lu$_2$Fe$_3$Si$_5$, recent measurements of specific heat \cite{Nakajima} and penetration depth \cite{Gordon} reported the evidence for two-gap superconductivity, similar to MgB$_2$ which is considered to be a two-gap $s$-wave superconductor.\cite{Choi} The Josephson effect suggested the spin-singlet superconductivity in Lu$_2$Fe$_3$Si$_5$.\cite{Noer} On the other hand, past experimental studies in Lu$_2$Fe$_3$Si$_5$ reported peculiar superconducting properties which are different from MgB$_2$: for instance, a power-law temperature dependence of specific heat below $T_c$,\cite{Vining} and a remarkable depression of $T_c$ by non-magnetic impurities.\cite{Xu2,Braun3} In addition, recent photoemission spectroscopy in the superconducting state observed the gap opening without distinct coherence peaks implying the nodal structure,\cite{Baba} in contrast to the two coherence peaks clearly observed in MgB$_2$.\cite{Tsuda} It should be noted that "cleanliness" in terms of the electron mean-free path is necessary and common conditions to the occurrence of the multigap and the non-$s$-wave (e.g., $p$- or $d$-wave) superconductivities, and thus these are co-occurrable in the "clean" system.\cite{Kasahara} In the multigap system, we should also take into account another possibility of the extended $s$-wave ($s_{\pm}$-wave) superconductivity in which the sign of the order parameter changes between the different Fermi sheets. This has recently been supposed as a possible pairing symmetry for the FeAs systems.\cite{Mazin} The recent and the past experimental reports in Lu$_2$Fe$_3$Si$_5$ require studies on verification of the sign reversal of the superconducting order parameter.

The effect of impurity scattering is sensitive to the phase of the superconducting gap function.\cite{Balatsky} The $s$-wave superconductivity is robust against non-magnetic impurities while strongly suppressed by magnetic impurities. On the contrary, the non-$s$-wave even-parity superconductivity, with the presence of nodes in the gap, is sensitive to both non-magnetic and magnetic impurities. The $s_{\pm}$-wave superconductivity, with the sign change of the order parameter between the different Fermi sheets, is expected to exhibit the impurity effects similar to the non-$s$-wave even-parity superconductivity.\cite{Golubov}

This paper reports study of non-magnetic and magnetic impurity effects on the superconductivity of Lu$_2$Fe$_3$Si$_5$ by small-amount substitution of non-magnetic Sc, Y, and magnetic Dy for Lu. Earlier, a brief account of magnetic susceptibility studies in the solid solutions (Lu$_{1-x}R_x$)$_2$Fe$_3$Si$_5$ ($R$ = Sc, Y, Dy-Tm) was reported in which $T_c$ was depressed with $R$ substitutions.\cite{Braun3} The present study particularly takes interest in the effect of disorder on the superconductivity in Lu$_2$Fe$_3$Si$_5$, and we study the correlation between $T_c$ and residual resistivity. We investigate structural, magnetic, and electrical properties of polycrystalline (Lu$_{1-x}R_x$)$_2$Fe$_3$Si$_5$ ($R$ = Sc, Y, and Dy). In addition, we investigate anisotropy of electrical resistivity in a high-purity Lu$_2$Fe$_3$Si$_5$ single crystal, motivations of which are described later with the results.

Polycrystals of (Lu$_{1-x}$Sc$_x$)$_2$Fe$_3$Si$_5$, (Lu$_{1-x}$Y$_x$)$_2$Fe$_3$Si$_5$ ($x$ = 0 - 0.07, and 1), and (Lu$_{1-x}$Dy$_x$)$_2$Fe$_3$Si$_5$ ($x$ = 0 - 0.05, and 1) were prepared by arc melting stoichiometric amounts of high-purity elements in Zr-gettered Ar atmosphere. To ensure the sample homogeneity, the arc melting was repeated with turning over the melted ingot for more than ten times. A high-purity single crystal of Lu$_2$Fe$_3$Si$_5$ was grown by the floating-zone method. The poly- and the single-crystalline samples were annealed at 1050Ž for 2 weeks. Powder X-ray diffraction patterns showed that all the samples crystallize in the Sc$_2$Fe$_3$Si$_5$-type structure without any additional peak. DC magnetic susceptibilities and electrical resistivities were measured by using the Quantum Design PPMS.

\begin{figure}[t]
\begin{center}
\includegraphics[scale=0.32]{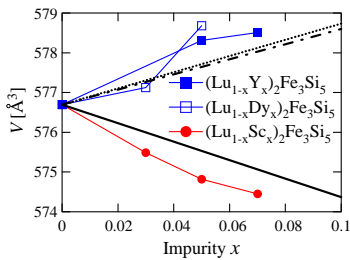}
\caption{\label{fig:Fig1} (Color online). The unit-cell volume of (Lu$_{1-x}$Sc$_x$)$_2$Fe$_3$Si$_5$, (Lu$_{1-x}$Y$_x$)$_2$Fe$_3$Si$_5$, and (Lu$_{1-x}$Dy$_x$)$_2$Fe$_3$Si$_5$ as a function of impurity concentration $x$. Solid, dotted, and dashed lines denote the Vegard's law in (Lu$_{1-x}$Sc$_x$)$_2$Fe$_3$Si$_5$, (Lu$_{1-x}$Y$_x$)$_2$Fe$_3$Si$_5$, and (Lu$_{1-x}$Dy$_x$)$_2$Fe$_3$Si$_5$, respectively.}
\end{center}
\end{figure}

Figure 1 depicts the unit-cell volume of (Lu$_{1-x}$Sc$_x$)$_2$Fe$_3$Si$_5$, (Lu$_{1-x}$Y$_x$)$_2$Fe$_3$Si$_5$, and (Lu$_{1-x}$Dy$_x$)$_2$Fe$_3$Si$_5$ as a function of impurity concentration $x$. The Vegard's law lines expected from the unit-cell volume of Lu$_2$Fe$_3$Si$_5$ (576.7 $\AA$), Sc$_2$Fe$_3$Si$_5$ (553.4 $\AA$), Y$_2$Fe$_3$Si$_5$ (597.1 $\AA$), and Dy$_2$Fe$_3$Si$_5$ (595.7 $\AA$) are also presented. It is evident that all the samples obey the Vegard's law: the unit-cell volume increases with $x$ in (Lu$_{1-x}$Y$_x$)$_2$Fe$_3$Si$_5$ and (Lu$_{1-x}$Dy$_x$)$_2$Fe$_3$Si$_5$, while decreases with $x$ in (Lu$_{1-x}$Sc$_x$)$_2$Fe$_3$Si$_5$. These results ensure that Y, Sc, and Dy atoms are properly introduced as impurities into the parent Lu$_2$Fe$_3$Si$_5$ phase with the Lu-site substitutions.

\begin{figure}[b]
\begin{center}
\includegraphics[scale=0.32]{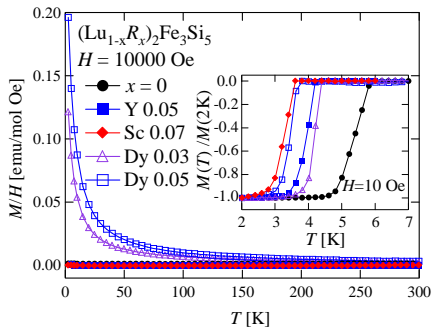}
\caption{\label{fig:Fig2} (Color online). The magnetic susceptibility of polycrystalline Lu$_2$Fe$_3$Si$_5$, (Lu$_{1-x}$Y$_x$)$_2$Fe$_3$Si$_5$ ($x$ = 0.05), (Lu$_{1-x}$Sc$_x$)$_2$Fe$_3$Si$_5$ ($x$ = 0.07), and (Lu$_{1-x}$Dy$_x$)$_2$Fe$_3$Si$_5$ ($x$ = 0.03 and 0.05) as a function of temperature with $H$ = 10000 Oe. Inset shows the superconducting transitions with $H$ = 10 Oe.}
\end{center} 
\end{figure}

Figure 2 depicts the magnetic susceptibility of the polycrystalline Lu$_2$Fe$_3$Si$_5$, (Lu$_{1-x}$Y$_x$)$_2$Fe$_3$Si$_5$ ($x$ = 0.05), (Lu$_{1-x}$Sc$_x$)$_2$Fe$_3$Si$_5$ ($x$ = 0.07), and (Lu$_{1-x}$Dy$_x$)$_2$Fe$_3$Si$_5$ ($x$ = 0.03 and 0.05) as a function of temperature with $H$ = 10000 Oe. (Lu$_{1-x}$Dy$_x$)$_2$Fe$_3$Si$_5$ exhibits the pronounced Curie tail due to the inclusion of the magnetic Dy atoms, in contrast to the non-magnetic behavior in Lu$_2$Fe$_3$Si$_5$, (Lu$_{1-x}$Y$_x$)$_2$Fe$_3$Si$_5$, and (Lu$_{1-x}$Sc$_x$)$_2$Fe$_3$Si$_5$. We here estimate the concentration of Dy atoms in the present (Lu$_{1-x}$Dy$_x$)$_2$Fe$_3$Si$_5$ from the Curie-Weiss behavior. The magnetic moment of Dy atom in Dy$_2$Fe$_3$Si$_5$ estimated from the Curie-Weiss behavior is $\mu$ = 10.4 $\mu_B$ which is close to the free-ion value ($\mu$ = 10.6 $\mu_B$). Using $\mu$ = 10.4 $\mu_B$, the Curie-Weiss analysis tells that 3.07 $\%$ and 4.92 $\%$ of Lu atoms are substituted by Dy atoms in the $x$ = 0.03 and 0.05 samples of (Lu$_{1-x}$Dy$_x$)$_2$Fe$_3$Si$_5$, respectively, ensuring that the Dy atoms are properly doped as magnetic impurities in these samples. The inset to Fig. 2 displays the low-temperature magnetic susceptibilities with $H$ = 10 Oe, exhibiting the diamagnetism due to the superconducting transition. For all the samples applied in the present study, the onset of the diamagnetism coincides with that of the zero-resistance transition, and we adopt these onset temperatures as $T_c$.

\begin{figure}[b]
\begin{center}
\includegraphics[scale=0.32]{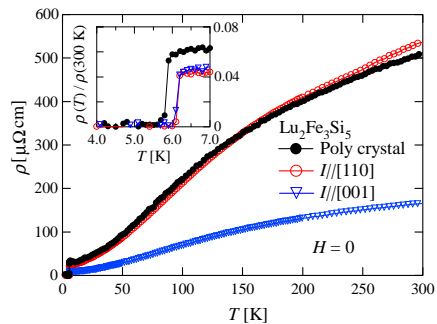}
\caption{\label{fig:Fig3} (Color online). The electrical resistivity of single-crystalline ($I$$\parallel$[001] and $I$$\parallel$[110]) and polycrystalline Lu$_2$Fe$_3$Si$_5$ as a function of temperature. Inset shows the low-temperature resistivities normalized to the values at 300 K.}
\end{center}
\end{figure}

The electrical resistivity of single- and poly-crystalline Lu$_2$Fe$_3$Si$_5$ is presented in Fig. 3 as a function of temperature. Superconducting transition occurs at $T_c$ = 6.1 K and 5.8 K in the single- and the poly-crystalline samples, respectively.  For the single crystal, we investigate the anisotropy of the resistivity with the current $I$ parallel and perpendicular to the crystal $c$-axis, $I \parallel$[001] and $I \parallel$[110], respectively. As shown in Fig. 3, the $c$-axis resistivity $\rho^c$ is less than one-third of the in-plane resistivity $\rho^{ab}$ in the whole temperature range, indicating the quasi one-dimensional conductivity in Lu$_2$Fe$_3$Si$_5$. The normal-state residual resistivities are $\rho^c_0$ = 7.0 $\mu \Omega$ cm and $\rho^{ab}_0$ = 22 $\mu \Omega$ cm, respectively. At 300 K, the polycrystalline resistivity $\rho^p$ exhibits an intermediate value between the single-crystalline $\rho^c$ and $\rho^{ab}$, $\rho^c(300 K)<\rho^p(300 K)<\rho^{ab}(300 K)$. The $\rho^p(300 K)$ is close to but smaller than $\rho^{ab}(300 K)$, indicating that $\rho^p$ is a weighted average of $\rho^c$ and $\rho^{ab}$ which dominantly picks up $\rho^{ab}$ as a component rather than $\rho^c$. As the temperature is lowered below $\sim$140 K, $\rho^p$ becomes slightly larger than $\rho^{ab}$. The normal-state residual resistivity of the poly crystal is $\rho^p_0$ = 30 $\mu \Omega$ cm which is larger than $\rho^c_0$ and $\rho^{ab}_0$, indicating that the poly crystal is "dirty" compared to the single crystal in terms of the electron mean-free path.

The inset to Fig. 3 shows the low-temperature resistivities $\rho^c$, $\rho^{ab}$, and $\rho^p$ normalized to the values at 300 K, $\rho(T)$/$\rho(300 K)$. It is evident that $\rho^c$ and $\rho^{ab}$ exhibit almost identical $\rho(T)$/$\rho(300 K)$: for the residual resistivities $\rho^c_0$ and $\rho^{ab}_0$, $\rho_0$/$\rho(300 K)$ = 0.04. Since $\rho(T)$/$\rho(300 K)$ cancels the contribution of the carrier density, and purely sees the variation of the electron mean-free path, the isotropy of $\rho(T)$/$\rho(300 K)$ in the single crystal indicates the isotropy of the electron mean-free path. Thus, it is ensured that the normalized resistivity $\rho(T)$/$\rho(300 K)$ is a good measure of the electron mean-free path regardless of single- and poly-crystals in Lu$_2$Fe$_3$Si$_5$. Similar to the "absolute" residual resistivities $\rho^p_0$, $\rho^c_0$, and $\rho^{ab}_0$, the normalized residual resistivity $\rho_0$/$\rho(300 K)$ in the inset to Fig. 3 tells that the polycrystalline Lu$_2$Fe$_3$Si$_5$ is "dirty" compared to the single crystal.

\begin{figure}[b]
\begin{center}
\includegraphics[scale=0.32]{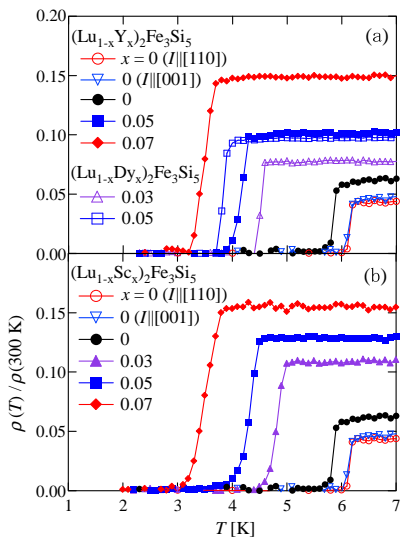}
\caption{\label{fig:Fig4} (Color online). Temperature dependence of the normalized resistivity $\rho(T)/\rho(300 K)$ for (a) (Lu$_{1-x}$Y$_x$)$_2$Fe$_3$Si$_5$ and (Lu$_{1-x}$Dy$_x$)$_2$Fe$_3$Si$_5$, and (b) (Lu$_{1-x}$Sc$_x$)$_2$Fe$_3$Si$_5$.}
\end{center} 
\end{figure}

On the basis of the isotropic electron mean-free path revealed by the single-crystalline resistivities, we now study the influence of disorder on the superconductivity in Lu$_2$Fe$_3$Si$_5$ by investigating the variation of $T_c$ with $\rho_0$/$\rho(300 K)$ in the polycrystalline samples. Fig. 4 (a) and (b) depict the normalized resistivity $\rho(T)$/$\rho(300 K)$ of non-magnetic (Lu$_{1-x}$Y$_x$)$_2$Fe$_3$Si$_5$ and (Lu$_{1-x}$Sc$_x$)$_2$Fe$_3$Si$_5$ as a function of temperature, respectively. Fig. 4 (a) also displays $\rho(T)$/$\rho(300 K)$ of magnetic (Lu$_{1-x}$Dy$_x$)$_2$Fe$_3$Si$_5$. It is noteworthy that the small amount of the Lu-site substitution in non-magnetic (Lu$_{1-x}$Y$_x$)$_2$Fe$_3$Si$_5$ and (Lu$_{1-x}$Sc$_x$)$_2$Fe$_3$Si$_5$ rapidly depresses $T_c$ with the systematic increase of the residual resistivity. Here, we would like to comment on the chemical pressure effect on $T_c$ in Lu$_2$Fe$_3$Si$_5$. The unit-cell volume variations in Fig. 1 tell that Y and Sc substitutions for Lu apply negative and positive chemical pressures, respectively. It is noted that $T_c$ of Y$_2$Fe$_3$Si$_5$ and Lu$_2$Fe$_3$Si$_5$ under hydrostatic pressure exhibits positive and negative pressure coefficients, $\frac{dT_c}{dp}>0$ and $\frac{dT_c}{dp}<0$, respectively.\cite{Segre} These $T_c$ variations imply that both the negative and the positive pressures might lower $T_c$ in Lu$_2$Fe$_3$Si$_5$. However, considering the difference of $T_c$ in Lu$_2$Fe$_3$Si$_5$ (6.1 K)$-$Y$_2$Fe$_3$Si$_5$ (2.6 K), and Lu$_2$Fe$_3$Si$_5$$-$Sc$_2$Fe$_3$Si$_5$ (4.6 K), the expected decrease of $T_c$ by the chemical pressure for (Lu$_{1-x}$Y$_x$)$_2$Fe$_3$Si$_5$ and (Lu$_{1-x}$Sc$_x$)$_2$Fe$_3$Si$_5$ at $x$ = 0.07 is $\Delta T_c = -0.25 $ K and $-0.1$ K, respectively. These values are much smaller than the $T_c$ depression of (Lu$_{1-x}$Y$_x$)$_2$Fe$_3$Si$_5$ and (Lu$_{1-x}$Sc$_x$)$_2$Fe$_3$Si$_5$ in Fig. 4, $\Delta T_c = -2.5$ K at $x$ = 0.07. Thus we conclude that the rapid $T_c$ depressions of Lu$_2$Fe$_3$Si$_5$ in Fig. 4 are dominated by the pair breaking by impurities. And the present results clearly indicate that {\it the introduction of disorder gives rise to the strong pair breaking in Lu$_2$Fe$_3$Si$_5$.} For magnetic (Lu$_{1-x}$Dy$_x$)$_2$Fe$_3$Si$_5$, $T_c$ is also steeply depressed with the Dy doping. Comparing the Dy- and Y-doped samples at $x$ = 0.05 which exhibit almost the same residual resistivities, $T_c$ = 3.8 K of (Lu$_{0.95}$Dy$_{0.05}$)$_2$Fe$_3$Si$_5$ is a little lower than $T_c$ = 4.2 K of (Lu$_{0.95}$Y$_{0.05}$)$_2$Fe$_3$Si$_5$. In (Lu$_{1-x}$Dy$_x$)$_2$Fe$_3$Si$_5$, as evident from Fig. 2, the Dy doping introduces the magnetic scattering potential. Thus the pair breaking in (Lu$_{1-x}$Dy$_x$)$_2$Fe$_3$Si$_5$ a little stronger than (Lu$_{1-x}$Y$_x$)$_2$Fe$_3$Si$_5$ is attributed to the magnetic scattering, which is compatible with the spin-singlet pairing in Lu$_2$Fe$_3$Si$_5$.\cite{Noer}

\begin{figure}[b]
\begin{center}
\includegraphics[scale=0.32]{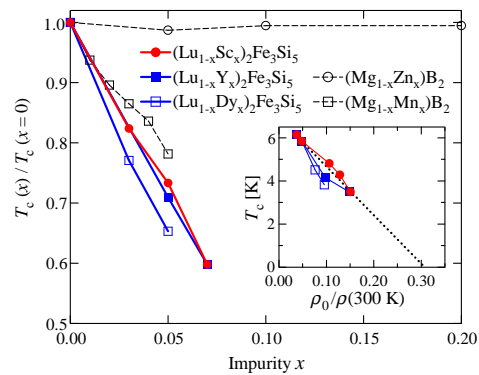}
\caption{\label{fig:Fig5} (Color online). The $T_c$ depression of Lu$_2$Fe$_3$Si$_5$ as a function of non-magnetic (Y, Sc) and magnetic (Dy) impurity concentrations $x$. Non-magnetic (Zn) and magnetic (Mn) impurity effects on $T_c$ in MgB$_2$ (Ref.~\cite{SXu}) are displayed for comparison. Inset shows the $T_c$ depression of Lu$_2$Fe$_3$Si$_5$ as a function of normalized residual resistivity $\rho_0$/$\rho$(300 K).}
\end{center} 
\end{figure}

Figure 5 displays non-magnetic and magnetic impurity effects on $T_c$ in Lu$_2$Fe$_3$Si$_5$ (this work) compared with MgB$_2$.\cite{SXu} For MgB$_2$, $T_c$ depression by non-magnetic impurities (Zn) is negligibly small while magnetic impurities (Mn) strongly depress $T_c$, indicative of the $s$-wave pairing. Lu$_2$Fe$_3$Si$_5$, on the other hand, exhibits strong $T_c$ depression with doping regardless of non-magnetic and magnetic impurities. As already mentioned in conjunction with Fig. 4, $T_c$ of Lu$_2$Fe$_3$Si$_5$ is rapidly depressed by non-magnetic impurities in accordance with the increase of residual resistivity. Such a disorder-sensitive superconductivity compellingly suggests {\it the sign reversal of the superconducting order parameter}.

In the sign-reversal order parameter, it is expected that the pair breaking by disorder results in vanishing of $T_c$ at a critical residual resistivity $\rho_0(0)$ in which the electron mean-free path $l_0$ is of the order of the superconducting coherence length $\xi_0$, $l_0 \simeq \xi_0$. The inset to Fig. 5 shows the $T_c$ depression of Lu$_2$Fe$_3$Si$_5$ as a function of normalized residual resistivity $\rho_0/\rho(300 K)$. The dotted line in this figure is a linear-fit to the experimental plots of non-magnetic (Lu$_{1-x}$Y$_x$)$_2$Fe$_3$Si$_5$ and (Lu$_{1-x}$Sc$_x$)$_2$Fe$_3$Si$_5$. Extrapolating this line to $T_c$ = 0 expects that the superconductivity disappears at $\rho_0(0)/\rho(300 K)\simeq$ 0.3. For the estimation of the critical residual resistivity $\rho_0(0)$, we assume that the temperature-dependent part of the resistivity, $\Delta\rho(T) = \rho(T)-\rho_0$, is independent of the small amount of the non-magnetic impurities. And we utilize $\Delta\rho(300 K)$ of the single-crystalline Lu$_2$Fe$_3$Si$_5$ in Fig. 3 for the $\rho_0(0)$ estimation: $c$-axis $\Delta\rho^c(300 K)$ = 158 $\mu \Omega$ cm, and in-plane $\Delta\rho^{ab}(300 K)$ = 513 $\mu \Omega$ cm, respectively. Using these $\Delta\rho(300 K)$ values, $\rho_0(0)/\rho(300 K)$ = $\rho_0(0)/[\rho_0(0)+\Delta\rho(300 K)]$ = 0.3 leads to the critical residual resistivities, $c$-axis $\rho^c_0(0)$ = 68 $\mu \Omega$ cm and in-plane $\rho^{ab}_0(0)$ = 220 $\mu \Omega$ cm, respectively.

Concerning the in-plane $\rho^{ab}_0(0)$ = 220 $\mu \Omega$ cm, we would like to roughly estimate the corresponding electron mean-free path $l_0^{ab}$ by using the formula, $l_0^{ab} = \frac{\hbar (3\pi^2)^{1/3}}{e^2n^{2/3}\rho^{ab}_0(0)}$. For Lu$_2$Fe$_3$Si$_5$, the in-plane Hall coefficient in low temperatures, $R_H \simeq$ 1.5 $\times$10$^{-9}$ m$^3$C$^{-1}$,\cite{Nakajima} leads to $\frac{1}{R_He} \simeq 4.2 \times 10^{27}$ m$^{-3}$. Substituting this $\frac{1}{R_He}$ value for the carrier density $n$ in the above $l_0^{ab}$ formula calculates $l_0^{ab} \simeq 22 \AA$. On the other hand, the upper critical field $\mu_0 H_{c2}(0) \simeq$ 13 T with $H \parallel c$ in Lu$_2$Fe$_3$Si$_5$ \cite{Nakajima, Gordon} calculates the in-plane coherence length $\xi^{ab}_0 \simeq 50 \AA$. These $l_0^{ab}$ and $\xi^{ab}_0$ are comparable within an order of magnitude, but $l_0^{ab}<\xi^{ab}_0$. We note here that the temperature-dependent $R_H$ in Lu$_2$Fe$_3$Si$_5$ is indicative of the multiband feature.\cite{Nakajima} In the multiband system, $\frac{1}{R_He}$ is no longer the correct expression for the carrier density, and  might become larger than the true carrier density when the contributions of electron and hole bands cancel each other in $R_H$.\cite{Hall} The smaller $l_0^{ab}$ than $\xi^{ab}_0$ might be attributed to the overestimation of $n$ due to the multiband feature.

The present study provides strong evidence for the sign reversal of the superconducting order parameter in the multigap structure in Lu$_2$Fe$_3$Si$_5$. However, the present study is insufficient to distinguish between the non-$s$-wave even-parity and the $s_{\pm}$-wave pairings. Further experiments which probe angle-resolved information, such as magnetothermal experiments with rotating magnetic field, and angle-resolved photoemission spectroscopy, should be performed to determine the superconducting gap structure of LuFe$_3$Si$_5$.

In summary, we studied the effect of non-magnetic and magnetic impurities on the superconductivity of Lu$_2$Fe$_3$Si$_5$ by small amount substitution of non-magnetic Y, Sc, and magnetic Dy for Lu. The rapid $T_c$ depression by non-magnetic impurities in accordance with the increase of residual resistivity reveals the disorder-sensitive superconductivity in Lu$_2$Fe$_3$Si$_5$, providing strong evidence for the sign reversal of the superconducting order parameter.

We thank T. Baba and Y. Nakajima for helpful comments. This work was partly supported by a Grant-in-Aid for Scientific Research from the Ministry of Education, Culture, Sports, Science and Technology of Japan.


\begin{thebibliography}{prb}
\bibitem{Hall} For instance, the Hall coefficient $R_H$ of a two-band system consisting of electron and hole bands is written as $R_H = (n_h\mu_h^2-n_e\mu_e^2)/[e(n_h\mu_h + n_e\mu_e)^2]$. Here $n_h$ ($n_e$) is the carrier density of the hole (electron) band, and $\mu_h$ ($\mu_e$) is the mobility of the hole (electron) band.
\end{thebibliography}
\end{document}